\documentclass[prl,twocolumn]{revtex4}

\usepackage{graphicx}
\usepackage{bbm}
\usepackage{amsmath}
\usepackage{amsfonts}
\usepackage{amssymb}

\begin{document}

\title{%\begin{flushright}\begin{tiny}UWthPh-2006-19\end{tiny}\end{flushright}
Nonlocality and entanglement in a strange system}
\author{Beatrix C. Hiesmayr}
\affiliation{Institute for Theoretical Physics, University of
Vienna, Boltzmanngasse 5, A-1090 Vienna, Austria}

%\date{Received: date / Revised version: date}

\begin{abstract}
We show that the relation between nonlocality and entanglement is
subtler than one naively expects. In order to do this we consider
the neutral kaon system---which is oscillating in time
(particle--antiparticle mixing) and decaying---and describe it as an
open quantum system. We consider a Bell--CHSH inequality and show a
novel violation for non--maximally entangled states. Considering the
change of purity and entanglement in time we find that, despite the
fact that only two degrees of freedom at a certain time can be
measured, the
neutral kaon system does not behave like a bipartite qubit system.
\end{abstract}

\maketitle

\section{Introduction}

In the last years many experiments have been performed which confirm
the peculiar predictions of the quantum theory, in particular the
existences of correlations which manifest themselves at two
different locations and fail to be explainable by local realistic
theories. As a powerful tool to detect nonlocality -- which ensures
secure communication, e.g. Ref.~\cite{GisinCrypto} -- are the famous
Bell inequalities. On the other hand there is currently a huge
business to develop entanglement measures using the quantum physics
tools: Hilbert space, observables and tensor products. A lot of
different measures have been proposed so far and only for bipartite
qubit systems the problem is satisfactorily solved. Yet, very
recently a third approach has been proposed: asking how huge can
non--local correlations be only assuming non-signaling (no
faster--than--light communication). Or differently stated: why is
Nature not more non--local than predicted by quantum theory, e.g.
Ref.~\cite{Popescu,ScaraniM}.

The purpose of this Letter is to shed light on the features of
nonlocality and entanglement for massive meson--antimeson systems,
in particular for the neutral kaon--antikaon system. The neutral
K-mesons or simply kaons are bound states of quarks and anti--quarks
or more precis the strangeness state $+1$, $K^0$, is composed of an
anti--strange quark and a down quark and the strangeness state $-1$,
$\bar K^0$, is composed of a strange and anti--down quark.

Interestingly, also for strange mesons entangled states can be
obtained, in analogy to the entangled spin up and down pairs, or H
and V polarized photon pairs. Such states are produced by $e^+
e^-$--colliders through the reaction $e^+ e^- \to \Phi \to K^0 \bar
K^0$, in particular at DA$\Phi$NE in Frascati, or they are produced
in $p\bar p$--collisions, like, e.g., at LEAR at CERN. There, a $K^0
\bar K^0$ pair is described at the time $t=0$ by the entangled
antisymmetric Bell state,
\begin{eqnarray}\label{antisymmetricBellstate}
|\psi^-\rangle&=&\frac{1}{\sqrt{2}}\big\lbrace
%|\Uparrow\rangle_l\otimes|\Downarrow\rangle_r-|\Downarrow\rangle_l\otimes|\Uparrow\rangle_r\big\rbrace\nonumber\\
%&=&\frac{1}{2}\big\lbrace
%|0\rangle_l\otimes|1\rangle_r-|1\rangle_l\otimes|0\rangle_r\big\rbrace\nonumber\\
%&=&\frac{1}{2}\big\lbrace
%|H\rangle_l\otimes|V\rangle_r-|V\rangle_l\otimes|H\rangle_r\big\rbrace\nonumber\\
%&=&\frac{1}{2}\big\lbrace
|K^0\rangle_l\otimes|\bar K^0\rangle_r-|\bar
K^0\rangle_l\otimes|K^0\rangle_r\big\rbrace\,,
%&=&\frac{1}{2}\big\lbrace
%|B^0\rangle_l\otimes|\bar B^0\rangle_r-|\bar B^0\rangle_l\otimes|B^0\rangle_r\big\rbrace\nonumber\\
%&=&\frac{1}{2}\big\lbrace
%|I\rangle_l\otimes|\Uparrow\rangle_r-|II\rangle_l\otimes|\Downarrow\rangle_r\big\rbrace\nonumber\\
%&=&\dots\;,
\end{eqnarray}
where $l$ denotes the particle moving to the left hand side and $r$
the particle moving to the right hand side.

Analogously to entangled photon systems for these systems Bell
inequalities can be derived, i.e. the most general Bell inequality
of the CHSH--type is given by (see Ref.~\cite{BH1})
\begin{eqnarray}\label{chsh}
\lefteqn{S_{k_n,k_m,k_{n'},k_{m'}}(t_1,t_2,t_3,t_4)=}\nonumber\\
&&\left|
E_{k_n,k_m}(t_1,t_2)-E_{k_n,k_{m'}}(t_1,t_3)\right|\nonumber\\
&&\qquad\quad+|E_{k_{n'},k_{m}}(t_4,t_2)+E_{k_{n'},k_{m'}}(t_4,t_3)|\leq
2\;.
\end{eqnarray}
Here Alice can choose on the kaon propagating to her left hand side
the ``quasi--spin'', i.e. a certain superposition of kaon and
antikaon $|k_n\rangle=\alpha |K^0\rangle+\beta |\bar K^0\rangle$,
and how long the kaon propagates, the time $t$. The same options are
given to Bob for the kaon propagating to the right hand side. As in
the usual photon setup, Alice and Bob can choose among two settings.
The expectation value $E_{k_n,k_m}(t_1,t_2)$ denotes then that Alice
chooses to measure the quasi--spin $k_n$ at time $t_1$ on the kaon
propagating to her side and Bob chooses to measure $k_m$ at time
$t_2$ on his kaon.

We notice now already that in the neutral kaon case we have more
options than in the photon case, we can vary in the quasi--spin
space or vary the detection times or both.

Let us first choose all times equal to zero and choose the
quasi--spin states $k_n=K_S, k_m=\bar K^0, k_{n'}=k_{m'}=K_1^0$
where $K_S$ is the short--lived eigenstate, one eigenstate of the
time evolution, and $K_1^0$ is the ${\cal CP}$ plus eigenstate. Here
$\cal{C}$ stands for charge conjugation and  $\cal{P}$ for parity.
The neutral kaon system is known for violating the combined
transformation ${\cal CP}$. In Ref.~\cite{BGH3} the authors show
that after optimizing the Bell inequality (BI) can be turned into
\begin{eqnarray}\label{BICP}
\delta\leq 0\;
\end{eqnarray}
where $\delta$ is the ${\cal CP}$ violating parameter in mixing.
Experimentally, $\delta$ corresponds to the leptonic asymmetry of
kaon decays which is measured to be $\delta=(3.27\pm0.12)\cdot
10^{-3}$. This value is in clear contradiction to the value required
by the BI above, i.e. by the premises of local realistic theories.
In this sense the violation of a symmetry in high energy physics is
connected to the violation of a Bell inequality, i.e. to
nonlocality\footnote{Or that is to say, for all times equal to zero
rather contextuality than nonlocality is tested. However, it has
been shown, Ref.~\cite{BH1} that also for $t\geq0$ the BI is
violated.}. This is clearly not the case for photons, they do not
violate the $\cal{CP}$ symmetry. Moreover, the ${\cal CP}$ violating
parameter is measured for single states, but it nevertheless gives
information on bipartite states.

Although the BI~(\ref{BICP}) is as loophole free as possible, the
probabilities or expectations values involved are not directly
measurable, because experimentally there is no way to distinguish
the short--lived state $K_S$ from the ${\cal CP}$ plus state $K_1^0$
directly.

In this work we want to investigate another choice for the Bell
inequality (\ref{chsh}), i.e. all quasispins are equal to those for
$\bar K^0$, and we are going to vary all four times
\begin{eqnarray}\label{BIstrangeness}
\lefteqn{S_{\bar K^0,\bar K^0,\bar K^0,\bar
K^0}(t_1,t_2,t_3,t_4)=}\nonumber\\
&&| E_{\bar K^0,\bar K^0}(t_1,t_2)-E_{\bar K^0,\bar
K^0}(t_1,t_3)|\nonumber\\
&&\qquad\quad+|E_{\bar K^0,\bar K^0}(t_4,t_2)+E_{\bar K^0,\bar
K^0}(t_4,t_3)|\leq 2\;.
\end{eqnarray}
This has the advantage that it can in principle be tested in
experiments: Alice and Bob insert at a certain distance from the
source (corresponding to the detection times) a piece of matter
forcing the incoming neutral kaon to react. Because the strong
interaction is strangeness conserving one obtains via the reaction
products if the incoming kaon is an antikaon or not. Note that
different to photons a $NO$ event does not mean that the incoming
kaon is a $K^0$ but also includes the case that it could have
decayed before. In principle, the strangeness content can also be
obtained via decay modes, but Alice and Bob have no way to force
their kaon to decay at a certain time, the decay mechanism is a
spontaneous event. However, a necessary condition to refute any
local realistic theory are \textit{active} measurements, i.e.
exerting the free will of the experimenter (for more details consult
Ref.~\cite{SBBGH}).

Our question is: \textbf{Can we violate the Bell--CHSH inequality
sensitive to strangeness (\ref{BIstrangeness}) for a certain initial
state and what is the maximum value?}

The first naive guess would be yes. In Refs.~\cite{BH1,SBBGH} the
authors studied the problem for the initial maximally entangled Bell
state, Eq.~(\ref{antisymmetricBellstate}), and they found that a
value greater than $2$ cannot be reached, i.e. one cannot refute any
local realistic theory. The reason is that the
particle--antiparticle oscillation is too slow compared to the decay
or vice versa, i.e., the ratio of oscillation to decay
$x=\frac{\Delta m}{\Gamma}$ is about $1$ for kaons and not $2$
necessary for a formal violation. A different view is that the decay
property acts as a kind of ``decoherence'', as we will show. From
decoherence studies we know that some states are more ``robust''
against a certain kind of decoherence than others, this leads to the
question if another maximally entangled Bell state or maybe a
different initial state would lead to a violation.

For that let us study first how single neutral kaons are handled via
open quantum systems and then proceed to entangled kaons and discuss
their entanglement and purity properties which gives us an insight
in the behavior of this strange two-state system. The formalism also
enables us to calculate the correct expectation values for arbitrary
initial states needed for the Bell inequality (\ref{BIstrangeness}).

Different kinds of Bell inequalities are discussed e.g. in
Refs.~\cite{Bramon3,Bramon4,Bramon5} and also decoherence models can
be investigated, e.g. Refs.~\cite{BDH1,CabanPolen2006}, and the
model proposed in the former reference has recently been tested via
experimental data Refs.~\cite{KLOEzeta2005,KLOEzeta2006}.

%%%%%%%%%%%%%%%%%%%%%%%%%%%%%%%%%%%%%%%%%%%%%%%%%%%%%%%%%%%%%%%%%%%%%%%%
\section{Open quantum formalism of decaying systems}

Neutral kaons are a decaying two--state system due to the
particle--antiparticle oscillation in time and are usually described
via an effective Schr\"odinger equation which we write in the
Liouville von Neumann form
\begin{eqnarray}\label{effectiveSchroedi}
\frac{d}{dt}\, \rho_{ss}&=&-i\, H_{eff}\, \rho_{ss} + i\,
\rho_{ss}\, H_{eff}^\dagger
\end{eqnarray}
where $\rho_{ss}$ is a $2\times 2$ matrix and the Hamiltonian
$H_{eff}$ is non-Hermitian. Using the Wigner-Weisskopf-approximation
the effective Hamilton can be calculated to be
$H_{eff}=H-\frac{i}{2}\Gamma$ where the mass matrix $H$ and the
decay matrix $\Gamma$ are both Hermitian and positive. Here the weak
interaction Hamiltonian responsible for decay is treated as a
perturbation and interactions between the final states are
neglected. This Wigner-Weisskopf approximation gives the exponential
time evolution of the two diagonal states of $H_{eff}$:
\begin{eqnarray}\label{exptime}
|K_{S/L}(t)\rangle &=& e^{-i m_{S/L} t} e^{-\frac{\Gamma_{S/L}}{2}
t} |K_{S/L}\rangle\,,
\end{eqnarray}
where $m_{S/L}$ and $\Gamma_{S/L}$ are the masses and decay
constants for the short/long--lived state $K_{S/L}$
($\Gamma_S\approx 600 \Gamma_L$; $\Delta
m=m_L-m_S\simeq\Gamma_S/2$). A kaon with strangeness $+1$ (kaon) or
$-1$ (antikaon) is a superposition of the two mass--eigenstates,
i.e. $|K^0\rangle\simeq\frac{1}{\sqrt{2}}
\{|K_S\rangle+|K_L\rangle\}$ and $|\bar
K^0\rangle\simeq\frac{1}{\sqrt{2}} \{-|K_S\rangle+|K_L\rangle\}$.
Here the small $\cal{CP}$ violation is safely neglected throughout
the Letter. What makes the neutral kaon systems so attractive for
many physical analyzes is the huge factor between the two decay
rates, i.e. $\Gamma_S\approx 600 \Gamma_L$, and that the strangeness
oscillation is $\Delta m=m_L-m_S\simeq\Gamma_S/2$.

Considering Eq.~(\ref{exptime}) we notice that the state is not
normalized for $t>0$. Indeed, we are not describing a system, for
$t>0$ a neutral kaon has a surviving and decaying component. In
Ref.~\cite{BGH4} the authors show that by enlarging the original
two--dimensional Hilbert space by at least two further dimensions
representing the decay product states, the non--Hermitian part of
$H_{eff}$ can be incorporated into the dissipator of the enlarged
space via a Lindblad operator. Thus the time evolution of neutral
kaons is described by an open quantum formalism, in particular by a
master equation \cite{Lindblad,GoriniKossakowskiSudarshan}
\begin{eqnarray}\label{masterequation}
\frac{d}{dt} \rho&=&-i [\cal{H},\rho]-\cal{D}[\rho]
\end{eqnarray}
where the dissipator under the assumption of complete positivity and
Markovian dynamics has the well known general form $ {\cal
D}[\rho]=\frac{1}{2}\sum_j ({\cal A}_j^\dagger{\cal
A}_j\rho+\rho{\cal A}_j^\dagger{\cal A}_j-2 {\cal A}_j\rho{\cal
A}_j^\dagger)$. The density matrix $\rho$ lives on
$\textbf{H}_{tot}=\textbf{H}_s\bigoplus\textbf{H}_f$ where $s$ and
$f$ denote ``surviving'' and ``decaying'' or ``final'' components,
and it has the following decomposition
\begin{equation}\label{rhotot}
\rho=\left(\begin{array}{cc} \rho_{ss}&\rho_{sf}\\
\rho_{sf}^\dagger&\rho_{ff}\end{array}\right)
\end{equation}
where $\rho_{ij}$ with $i,j=s,f$ denote $2\times 2$ matrices. The
Hamiltonian $\cal{H}$ is the Hamiltonian $H$ of the effective
Hamiltonian $H_{eff}$ extended to the total Hilbert space
$\textbf{H}_{tot}$, and $\Gamma$ of $H_{eff}$ defines a Lindblad
operator by $\Gamma=A^\dagger A$, i.e.
\begin{eqnarray*}
{\cal H}=\left(\begin{array}{cc} H&0\\
0&0\end{array}\right)\;,\; {\cal A}=\left(\begin{array}{cc} 0&0\\
A&0\end{array}\right)\quad\textrm{with}\quad A:
\textbf{H}_s\rightarrow \textbf{H}_f\,.
\end{eqnarray*}
Rewriting the master equation for $\rho$, Eq.~(\ref{rhotot}), on
$\textbf{H}_{tot}$
\begin{eqnarray}
\dot{\rho}_{ss}&=&-i[H,\rho_{ss}]-\frac{1}{2}\,\lbrace A^\dagger
A, \rho_{ss}\rbrace\;,\\
\dot{\rho}_{sf}&=&-i H \rho_{sf}-\frac{1}{2}\, A^\dagger
A\, \rho_{sf}\;,\\
\label{rhoff} \dot{\rho}_{ff}&=&A\,\rho_{ss}\,A^\dagger\,,
\end{eqnarray}
we notice that the master equation describes the original effective
Schr\"odinger equation (\ref{effectiveSchroedi}) but with properly
normalized states, Ref.~\cite{BGH4}. By construction the time
evolution of $\rho_{ss}$ is independent of $\rho_{sf}, \rho_{fs}$
and $\rho_{ff}$. Further $\rho_{sf}$ and $\rho_{fs}$ completely
decouple from $\rho_{ss}$ and thus can without loss of generality be
chosen to be zero, they are not physical and can never be measured.
With the initial condition $\rho_{ff}(0)=0$ the time evolution is
\textit{solely} determined by $\rho_{ss}$---as expected for a
spontaneous decay process---and is formally given by integrating
Eq.~(\ref{rhoff}).

%%%%%%%%%%%%%%%%%%%%%%%%%%%%%%%%%%%%%%%%%%%%%%%%%%%%%%%%%%%%%%%%%%%%%%%%
\section{Time evolution of single kaons}

Without loss of generality the initial state can be chosen in the
mass eigenstate basis $\{K_S,K_L\}$. The formal solution of
Eq.(\ref{masterequation}) ($\Gamma=\frac{1}{2}(\Gamma_S+\Gamma_L)$
and the numbers
$\rho_{SS}+\rho_{LL}=1$) is %\begin{widetext}
\begin{eqnarray}\label{densitysingle}
\rho(t)=\left(\begin{array}{cccc} e^{-\Gamma_S t} \rho_{SS}& e^{-i
\Delta m t-\Gamma t} \rho_{SL}&0&0\\
e^{i \Delta m t-\Gamma t} \rho_{SL}^*&e^{-\Gamma_L t} \rho_{LL}&0&0\\
0&0&F_L \rho_{LL}&X^*\\
0&0&X&F_S \rho_{SS}
\end{array}\right)\,.\nonumber\\
\end{eqnarray}
%\end{widetext}
with $F_{S/L}=1-e^{-\Gamma_{S/L} t}$ and $X=\frac{\sqrt{\Gamma_S
\Gamma_L}}{-i \Delta m-\Gamma}(1-e^{-i \Delta m t-\Gamma
t})\rho_{SL}$. Clearly, we have $Tr\rho(t)=1$ and the decay is
caused by the environment (treating the neutral kaon in QFT
formalism, the decay would be caused by the QCD vacuum). The
surviving part of the single kaon evolving in time is represented by
the upper $2\times 2$ block matrix $\rho_{ss}$, the lower one by the
decaying part $\rho_{ff}$.

\begin{figure*}
\center{(a)\includegraphics[width=200pt,keepaspectratio=true]{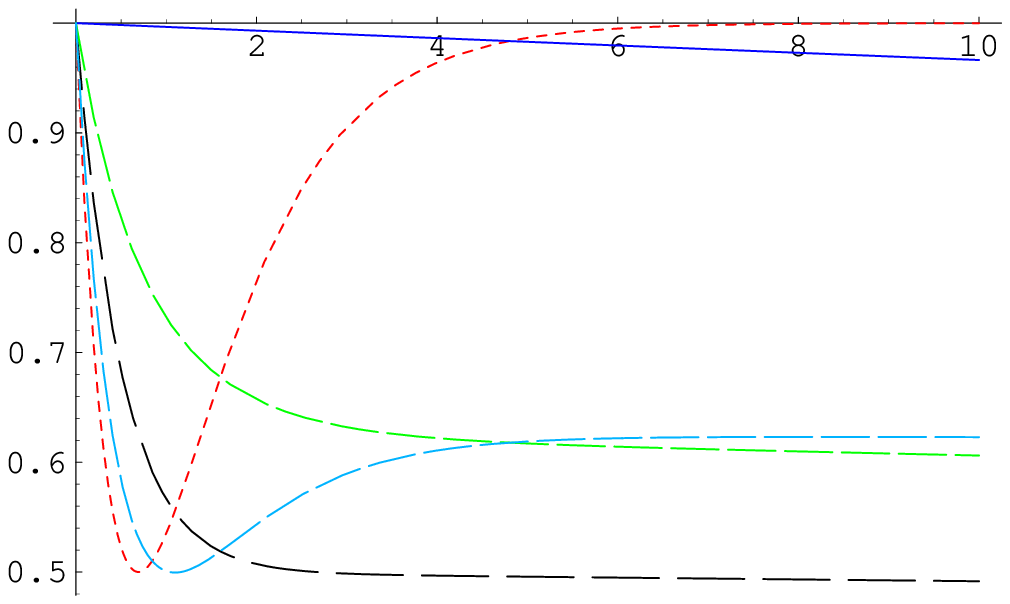}$\qquad\qquad$
(b)\includegraphics[width=200pt,keepaspectratio=true]{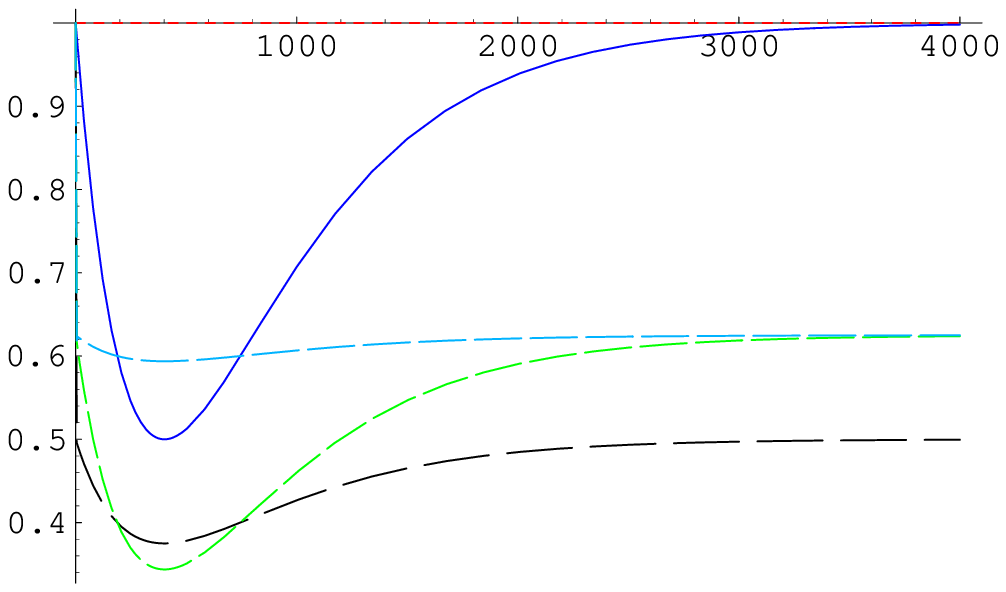}}
\caption{(Color online) Here the purity $Tr\rho(t)^2$ for single
kaons (initially pure) for short (a) and longer (b) time scales is
shown (units in $1/\Gamma_S$): (red (dashed): $K_S$; blue (solid):
$K_L$; black (long dashed) $K^0$ or $\bar K^0$; green: ${\tiny 1/2}
|K_S\rangle+\sqrt{3/4}|K_L\rangle$; light blue:
$\sqrt{3/4}|K_S\rangle+1/2|K_L\rangle$).}\label{puritysingle}
\end{figure*}

%XXX\begin{tiny}The off diagonal elements of $\rho_{ff}$ are not
%physical, they can in no way be measured, thus we can define them to
%be any value, e.g. for simplicity zero. The maximum value of the
%formal integration result $|\frac{\sqrt{\Gamma_S \Gamma_L}}{-i
%\Delta m-\Gamma}|(1-e^{i \Delta m t-\Gamma t})|$ is $\approx 0.06$,
%but is not related to $CP$ violation, which is not considered in
%this Letter (TRUE? Can we always add constants such that whole state
%always pure?).\end{tiny}XXX

Only properties of the surviving components can be measured. E.g. by
a piece of matter an incoming beam is forced to react with the
matter via the strong interaction (which is strangeness conserving).
If a reaction which can only be caused by a $\bar K^0$ is detected,
one records a yes--event (Y). If no $\bar K^0$ is detected a
no--event (N) is recorded (including a $K^0$ or a decay event). Then
matter acts in the very same manner as an ordinary polarisator for
photons. Note that an experimenter can \textit{actively} choose the
initial state (up to experimental realization), the kind of detector
(experimentally very limited) and where to place the detector, i.e.
how much ``decoherence'' the system undergoes, whereas the kind of
``decoherence'' is given by Nature. Note that this ``decoherence''
is fundamentally different from that in other quantum systems which
are stable, there the kind of decoherence depends on the
environment, for kaons it is intrinsic to the system.

Consequently, an operator $P$ projecting onto the states $\rho_{ss}$
gives the two probabilities, for $Y$ or $N$, that a certain state is
detected at time $t$:
%\begin{eqnarray}
%\left(\begin{array}{cc}
%P&0\\
%0&0
%\end{array}\right)\quad\textrm{and}\quad\left(\begin{array}{cc}
%\mathbbm{1}-P&0\\
%0&\mathbbm{1}
%\end{array}\right)\;.
%\end{eqnarray}
%and thus we can derive two different probabilities. The kaon was
%projected in a certain state at time $t$ ($Y$) or not ($N$):
\begin{eqnarray*}
Prob(Y,t)&=&Tr(\left(\begin{array}{cc}
P&0\\
0&0
\end{array}\right)\rho(t))=Tr(P\rho_{ss}(t))\quad\textrm{and}\nonumber\\
Prob(N,t)&=&Tr(\left(\begin{array}{cc}
\mathbbm{1}-P&0\\
0&\mathbbm{1}
\end{array}\right)\rho(t))\nonumber\\
&=&Tr((\mathbbm{1}-P)\rho_{ss}(t))+Tr(\rho_{ff}(t))\nonumber\\
&=&1-Tr(P\rho_{ss}(t))\;.
\end{eqnarray*}
Consequently, the expectation value becomes
$E_P(t)=Prob(Y,t)-Prob(N,t)=2\, Tr(P\rho_{ss}(t))-1$ and is
\textbf{solely} determined by the surviving component $\rho_{ss}$!

We considered all possible projectors and the $\rho_{ff}$ enters in
the probabilities only via the trace, thus it is clear that the off
diagonal elements of $\rho_{ff}$ are not relevant for any
probability we may derive. This leaves a certain ambiguity in
defining the decaying components and therefore purity and
entanglement. We choose the off diagonal elements of $\rho_{ff}$ in
Eq.~(\ref{densitysingle}) equal to zero because they give the lowest
purity values.

Let us now consider the change of the properties of the state
$\rho(t)$ with time by considering the purity defined by
\begin{eqnarray*}
\lefteqn{Tr\rho(t)^2=Tr(\rho_{ss}(t)^2)+Tr(\rho_{ff}(t)^2)}\nonumber\\
&=&Tr\rho_{ss}(t)^2+(Tr\rho_{ff}(t))^2\nonumber\\
&=&Tr\rho_{ss}(t)^2+(1-Tr\rho_{ss}(t))^2\nonumber\\
&=&\rho_{SS}^2(1-2 e^{-\Gamma_S t}+2 e^{-2\Gamma_S
t})\nonumber\\
&& +\rho_{LL}^2(1-2 e^{-\Gamma_L t}+2 e^{-2\Gamma_L t})+2
|\rho_{SL}|^2 e^{-2 \Gamma t}\,.
\end{eqnarray*}
Note that the second equality sign is only true if the off diagonal
elements of $\rho_{ff}$ vanish. Otherwise we would add an additional
in general time dependent factor to the definition of the purity
(for the formal integration of the order $10^{-2}$). Again our
definition of the purity is only depending on the surviving
components. Starting with an arbitrary initial pure state we see
that the decay ability of the system leads to a decrease in purity
for $t>0$. For $K_S$ or $K_L$ the purity returns to $1$ for
$t\rightarrow \infty$ depending on the decay constants, see
Fig.~\ref{puritysingle}. After a time $t/\tau_{S/L}=\ln 2$ the
minimal purity of $0.5$ of a usual qubit system described by a
$2\times 2$ density matrix (trace state) is reached. For other
superpositions the purity oscillates to a certain final purity which
$\not=1$. For an initial $K^0$ or $\bar K^0$ we reach the minimal
purity of $0.375$ at time $t/\tau_S=401.881$, i.e. about $2/3$ of
the lifetime of the long--lived state. This is much lower than the
purity of a qubit system. Indeed, this decaying system---where only
two degrees of freedom can be measured---behaves as regards the
purity properties as a system with more degrees of freedom, neutral
kaons are more like a double slit evolving in time, see
Ref.~\cite{SBGH2}. Clearly, we could renormalize the purity by
choosing appropriate off diagonal elements of $\rho_{ff}$.

\begin{figure*}
\center{{
(a)}\includegraphics[width=225pt,keepaspectratio=true]{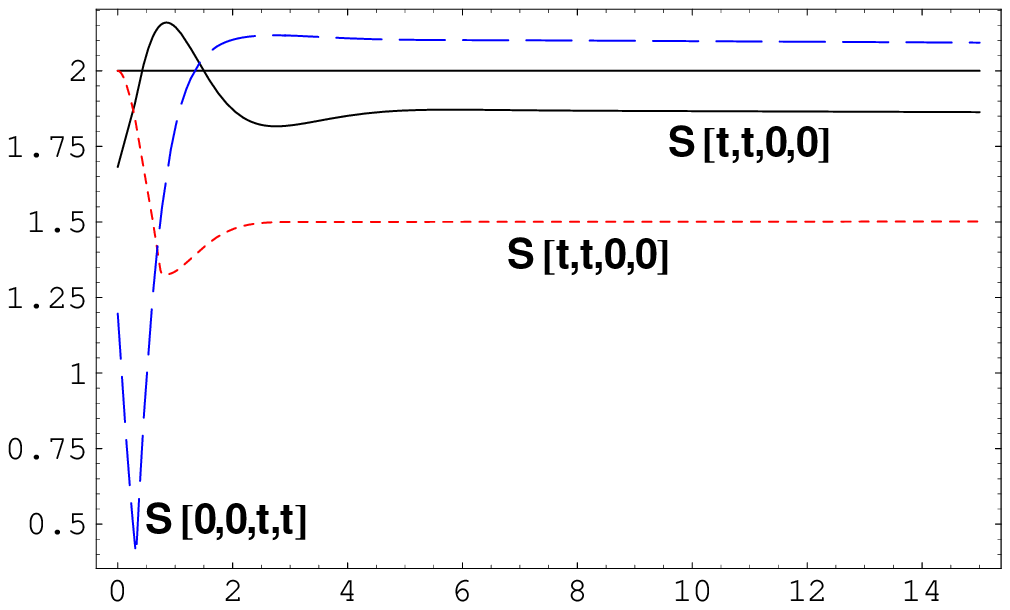}
{
(b)}\includegraphics[width=225pt,keepaspectratio=true]{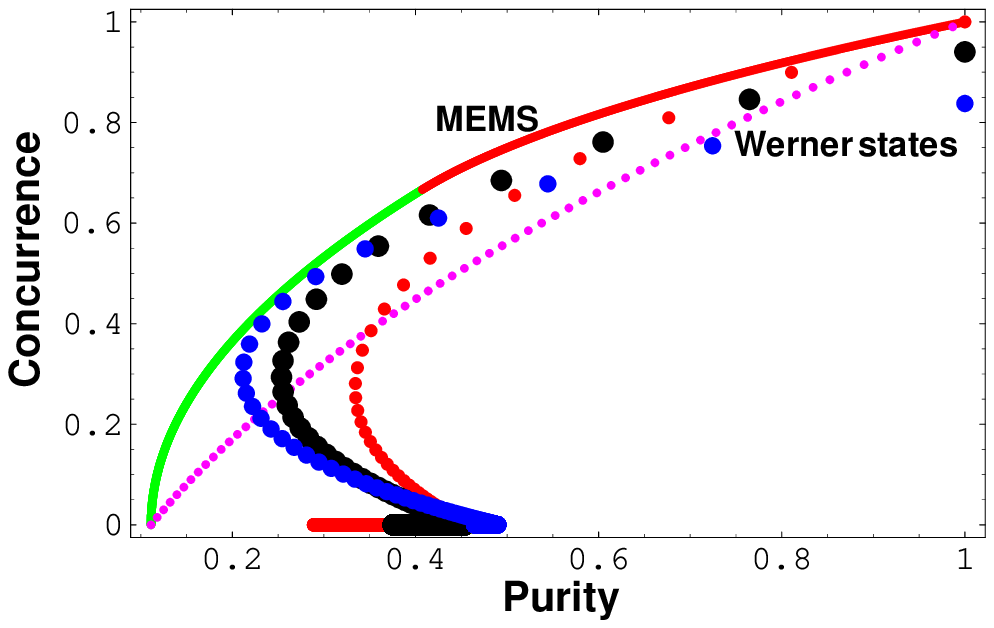}\\
{
(c)}\includegraphics[width=225pt,keepaspectratio=true]{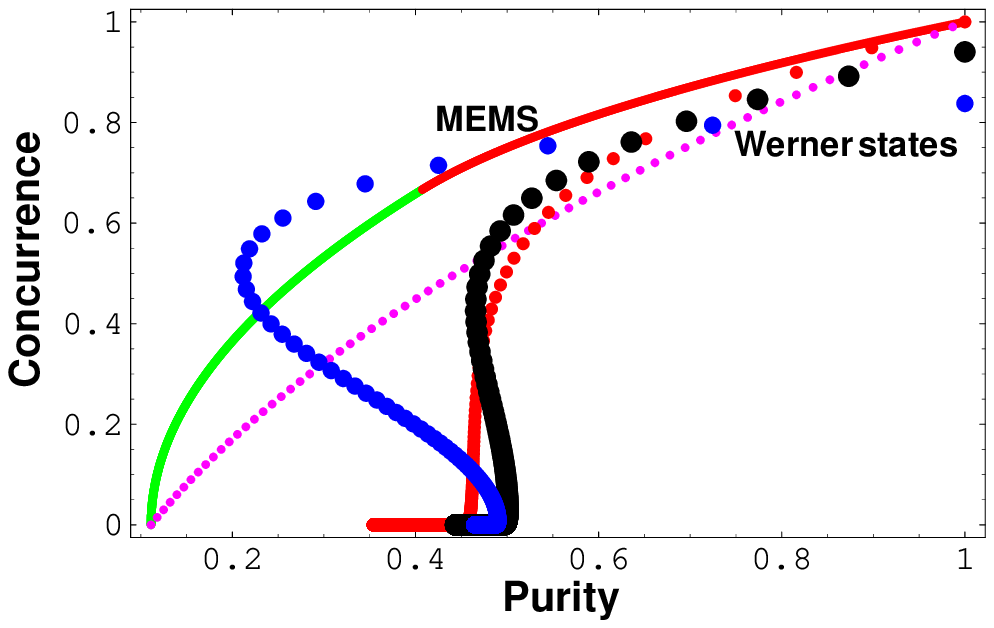}
{
(d)}\includegraphics[width=225pt,keepaspectratio=true]{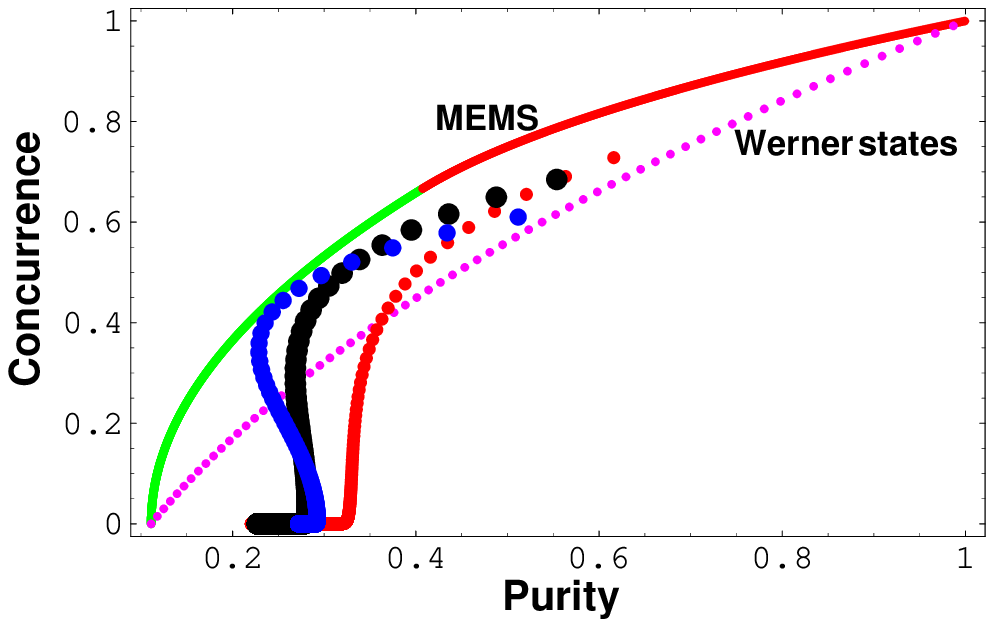}}\caption{(Colour
online) In Fig.(a) is shown the timedependent $S$--function for
$\phi^+$ (\textit{dashed, red}), $\xi$ (\textit{long dashed, blue})
and $\chi$ (\textit{solid, black}) (time in units of
$[\frac{\Gamma_S}{\Delta m}]$). For $\xi$ the violation exists up to
$1/\Gamma_L$. In Fig. (b)-(d) a purity versus concurrence diagram is
drawn (purity normalized $(d Tr\rho^2-1)/(d-1)$ with $d=4$ for
bipartite qubits and $d=16$ for bipartite kaons). The limiting curve
represents the maximally entangled mixed bipartite qubit states
(MEMS) \cite{MEMS} and the nearly linear curve (\textit{dashed,
purple}) the Werner states for bipartite qubits. The dots are drawn
for different initial states and the time proceeds from $0$ to $100$
with a step width of $0.05$ (units as above). The smallest dots
(\textit{red}) for $\phi^+$, next to smallest dots (\textit{blue})
$\xi$ and the biggest dots (\textit{black}) are for $\chi$. In
Fig.(b) is shown the change in purity and concurrence for
$t_l=t_r=t$, in (c) for $t_l=0; t_r=t$ and in (d) for $t_l=0.3;
t_r=t$.}\label{CversusmixedS}
\end{figure*}

The minimal purity which can be reached for this decaying system is
$0.333068$ obtained by an initially mixed state ($\rho_{SS}=2/3;
\rho_{SL}=0; t/\tau_S=0.694012$), and is thus greater than $0.25$,
the minimal value for a $4\times 4$ density matrix (trace state).
%Though it is slightly lower than $1/3$ (qutrit system).
Note that in general it depends on the ratio between
$\Gamma_S/\Gamma_L$ and is therefore intrinsic to the described
meson system.\\
%As also
%visualized in Fig.~\ref{puritysingle} the greater the weight of the
%component $K_L$ is in an initial superposition the more mixed the
%state gets during it's time evolution. However, the decrease in the
%purity is greater for short times if the weight of $K_S$ is greater.

%%%%%%%%%%%%%%%%%%%%%%%%%%%%%%%%%%%%%%%%%%%%%%%%%%%%%%%%%%%%%%%%%%%%%
\section{The time evolution for two kaons}

Any density matrix of a single kaon evolving in time,
Eq.~(\ref{densitysingle}), can be decomposed in the following way $$
\rho(t)=\sum_{nm} f_{nm}(t) \rho_{nm} \;|n\rangle\langle m|\;. $$
Clearly, for two kaons in a product state we have $$
\sigma(t)=\sum_{nmlk} f_{nm}(t) f_{lk}(t) \rho_{nm}\rho_{lk}
\;|n\rangle\langle m|\otimes|l\rangle\langle k|\,,$$ and,
consequently, any two--kaon state is then given by
\begin{eqnarray}
\sigma(t)&=&\sum_{nmlk} f_{nm}(t) f_{lk}(t) \sigma_{nmlk}
\;|n\rangle\langle m|\otimes|l\rangle\langle k|\;,
\end{eqnarray}
where the time dependent weights can be assumed to factorize. In
order to do this one has to prove that the projectors commute with
the generators of the time evolution under the trace (this was
proven in a different formulation in Ref.~\cite{ThesisTrixi}). We
can even define a two--particle density matrix depending on the two
different times representing the times when the two kaons are
measured, i.e.
\begin{eqnarray*}
\sigma(t_l,t_r)&=&diag\lbrace\sigma_{ssss}(t_l,t_r),\sigma_{ssff}(t_l,t_r),\nonumber\\
&& \sigma_{ffss}(t_l,t_r),\sigma_{ffff}(t_l,t_r)\rbrace\,,
\end{eqnarray*}
where $\sigma_{iijj}$ are $4\times 4$ matrices.
%with (appendix??)
%\begin{small}
%\begin{eqnarray}
%\sigma_{ssss}(t_l,t_r)&=&\left(\begin{array}{cccc} e^{-\Gamma_S
%t_l-\Gamma_S t_r}\sigma_{11}&e^{-\Gamma_S t_l}e^{-i \Delta m
%t_r-\Gamma t_r}\sigma_{12}
%&e^{-i \Delta m t_l-\Gamma t_l} e^{-\Gamma_S t}\sigma_{13}&e^{(-i \Delta m-\Gamma)(t_l+t_r)}\sigma_{14}\\
%e^{-\Gamma_S t_l}e^{-i \Delta m t_r-\Gamma
%t_r}\sigma_{12}^*&e^{-\Gamma_S
%t_l}e^{-\Gamma_L t_r}\sigma_{22}& e^{-i \Delta m (t_l-t_r)-\Gamma (t_l+t_r)}\sigma_{23}&e^{-i\Delta m t_l-\Gamma t_l-\Gamma_L t}\sigma_{24}\\
%\sigma_{13}^*&\sigma_{23}^*&\sigma_{33}&\sigma_{34}\\
%\sigma_{14}^*&\sigma_{24}^*&\sigma_{34}^*&\sigma_{44}
%\end{array}\right)\;,\nonumber\\
%\sigma_{ssff}&=&\left(\begin{array}{cccc}
%\sigma_{11}&0&\sigma_{13}&0\\
%0&\sigma_{22}&0&\sigma_{24}\\
%\sigma_{13}^*&0&\sigma_{33}&0\\
%0&\sigma_{24}^*&0&\sigma_{44}
%\end{array}\right)\;,\nonumber\\
%\sigma_{ffss}&=&\left(\begin{array}{cccc}
%\sigma_{11}&\sigma_{12}&0&0\\
%\sigma_{12}^*&\sigma_{22}&0&0\\
%0&0&\sigma_{33}&\sigma_{34}\\
%0&0&\sigma_{34}^*&\sigma_{44}
%\end{array}\right)\;,\nonumber\\
%\sigma_{ffff}&=&\left(\begin{array}{cccc}
%\sigma_{11}&0&0&0\\
%0&\sigma_{22}&0&0\\
%0&0&\sigma_{33}&0\\
%0&0&0&\sigma_{44}
%\end{array}\right)\;.\end{eqnarray}
%\end{small}
%We are now interested in the entanglement of such a density matrix.
As a measure of entanglement we want to consider the entanglement of
formation which is defined by $\mathcal{E}o\mathcal{F}(\rho)=min_i
\sum_i p_i S(Tr_l( |\psi_i\rangle\langle\psi_i|))$ where $S$ is the
von Neumann entropy, the trace is taken over one subsystem (left or
right) and $\psi_i$ are the pure state decompositions of $\rho$. A
necessary criterion for entanglement is that the matrix under
partial transposition ($PT$) has at least one negative eigenvalue.
Only for bipartite two--level systems $PT$ is also sufficient for
detecting all entangled states. For the density matrix under
investigation $PT$ acts in the following way
\begin{eqnarray*}
PT[\sigma(t_l,t_r)]&=&diag\lbrace PT[\sigma_{ssss}(t_l,t_r)],PT[\sigma_{ssff}(t_l,t_r)],\nonumber\\
&& PT[\sigma_{ffss}(t_l,t_r)],PT[\sigma_{ffff}(t_l,t_r)]\rbrace\;.
\end{eqnarray*}
The surviving--surviving block $\sigma_{ssss}$ can lead to negative
eigenvalues, i.e. can be entangled, while the eigenvalues of the
other blocks cannot become negative due to the vanishing off
diagonal elements, the eigenvalues remain unchanged under $PT$. Thus
whether the state under investigation is entangled depends only on
$\sigma_{ssss}$. For $4\times 4$ matrices entanglement of formation
is an increasing function of the computable concurrence
$\mathcal{C}$, found by Hill and Wootters
Ref.~\cite{WoottersHill}\footnote{For higher dimensions no
computable function of entanglement of formation is known.}. Thus we
can measure entanglement by the concurrence of $\sigma_{ssss}$.

To compute concurrence one defines the flipped matrix
$\tilde\sigma_{ssss}=(\sigma_y\otimes\sigma_y)\sigma_{ssss}^*(\sigma_y\otimes\sigma_y)$
where $\sigma_y$ is the $y$--Pauli matrix and the complex
conjugation is taken in the $K_S K_L$ basis. The concurrence is then
given by the formula
$\mathcal{C}=max\{0,\lambda_1-\lambda_2-\lambda_3-\lambda_4\}$ where
the $\lambda_i$'s are the square roots of the eigenvalues, in
decreasing order, of the matrix $\sigma_{ssss}\tilde\sigma_{ssss}$.

Let us now consider a general pure state at $t=0$ (with
$r_1^2+r_2^2+r_3^2+r_4^2=1$ ; $\otimes$ omitted)
\begin{eqnarray}\label{generalinitialstate}
|\psi(0)\rangle &=& r_1 e^{i \phi_1} |K_S\rangle|K_S\rangle +r_2
e^{i \phi_2} |K_S\rangle|K_L\rangle\nonumber\\
&& +r_3 e^{i \phi_3} |K_L\rangle|K_S\rangle+r_4 e^{i \phi_4}
|K_L\rangle|K_L\rangle\;.
\end{eqnarray}
Alice and Bob perform their measurements at certain times $t_l,
t_r$, respectively. For a general initial pure state the concurrence
is derived to be $$\mathcal{C}(\sigma_{ssss}(t_l,t_r))=2\; |r_1 r_4
e^{i \phi_1+i\phi_4}-r_2 r_3 e^{i\phi_2+i\phi_3}|\; e^{-\Gamma
(t_l+t_r)}\;.$$ It is simply the concurrence of the initial pure
state multiplied by the time depending damping factor. For one time
equal to zero the decrease in entanglement is lowest.

%\begin{figure} \center{
%\includegraphics[width=105pt, keepaspectratio=true]{Sviolation.EPS}
%}\caption{(Colour online) Here the function $S_{K^0,K^0}(t,t,0,0)$
%for an initial $|\phi^+\rangle=\frac{1}{\sqrt{2}}
%\{|K_S\rangle|K_S\rangle+|K_L\rangle|K_L\rangle\}$ (blue dashed
%curve) and the function $S_{K^0,K^0}(0,t,0,t)$ for $r_1=0.776679;
%r_2=-0.232662; r_3=0.232663; t_1=t_3=0; t_2=1.53014; t_4=1.53014;
%\phi_1=-0.776486; \phi_2=-0.346577; \phi_3=2.79502; \phi_4=1.68703$
%is plotted. The maximum value of the $S$-function is $2.00304$ and
%$2.18406$, respectively.}
% \label{Sviolation}
%\end{figure}
%
%For the two--particle state we obtain four probabilities
%\begin{eqnarray}
%Prob(Y,t_l;Y,t_r)&=&Tr(P_l\otimes P_r\sigma_{ssss})\;,\nonumber\\
%Prob(Y,t_l;N,t_r)&=&Tr((1-P_l)\otimes\mathbbm{1}(\sigma_{ssss}+\sigma_{ssff}))\nonumber\\
%Prob(N,t_l;Y,t_r)&=&Tr(\mathbbm{1}\otimes(1-P_r)(\sigma_{ssss}+\sigma_{ffss}))\nonumber\\
%Prob(N,t_l;N,t_r)&=&1+Tr(P_l\otimes
%P_r\sigma_{ssss})\nonumber\\
%\lefteqn{-Tr((1-P_l)\otimes\mathbbm{1}\sigma_{ssff})-(1-P_r)(\sigma_{ffss})}\nonumber
%\end{eqnarray}
%and the expectation value derives to
%$E_{P_l,P_r}(t_l,t_r)=%Prob(Y, t_l; Y, t_r)+Prob(N, t_l; N,
%%t_r)-Prob(Y, t_l; N, t_r)-Prob(N, t_l; Y, t_r)\nonumber\\
%%&=&
%-1+2(Prob(Y, t_l; Y, t_r)+Prob(N, t_l; N, t_r))$ where we used the
%fact that the sum of all probabilities gives one.
We choose as projectors  $P_{r,l}=|\bar K^0\rangle\langle \bar
K^0|$, and the expectation value becomes after a cumbersome
calculation
%\begin{widetext}
\begin{eqnarray}\label{expectationvalue}
\lefteqn{E_{\bar K^0,\bar K^0}(t_l,t_r)\; =\; 1+ r_1^2\,
e^{-\Gamma_S (t_l+t_r)}+r_2^2\, e^{-\Gamma_S t_l-\Gamma_L
t_r}}\nonumber\\
&&+r_3^2\, e^{-\Gamma_L
t_l-\Gamma_S t_r}+r_4^2\, e^{-\Gamma_L (t_l+t_r)}\nonumber\\
&&- r_1^2\, (e^{-\Gamma_S t_l}+ e^{-\Gamma_S t_r})-r_2^2\,
(e^{-\Gamma_S t_l}+ e^{-\Gamma_L t_r})\nonumber\\
&& - r_3^2\, (e^{-\Gamma_L t_l}+ e^{-\Gamma_S t_r}) - r_4^2\,
(e^{-\Gamma_L t_l}+ e^{-\Gamma_L
t_r})\nonumber\\
&&+2\,  r_1 r_2\, (1-e^{-\Gamma_S t_l}) \cos(\Delta m
t_r+\phi_1-\phi_2)\, e^{-\Gamma t_r}\nonumber\\
&&+2\, r_1 r_3\, \cos(\Delta m
t_l+\phi_1-\phi_3)\, e^{-\Gamma t_l}\,(1-e^{-\Gamma_S t_r})\nonumber\\
&&+ 2\,  r_2 r_4\,  \cos(\Delta m t_l+\phi_2-\phi_4)\, e^{-\Gamma
t_l}\, (1-e^{-\Gamma_L t_r})\nonumber\\
&&+2\, r_3 r_4\, (1-e^{-\Gamma_L t_l})\,\cos(\Delta m
t_r+\phi_3-\phi_4)\,
 e^{-\Gamma t_r}\nonumber\\
&&+ 2\,  r_1 r_4 \cos(\Delta m (t_l+t_r)+\phi_1-\phi_4)\, e^{-\Gamma
(t_l+t_r)}\nonumber\\
&&+2\, r_2 r_3\, \cos(\Delta m (t_l-t_r)+\phi_2-\phi_3)\, e^{-\Gamma
(t_l+t_r)}\;.
\end{eqnarray}
%\end{widetext}
We notice that for any initial state one always has damping
functions from the decay property in this system different from
other two--state systems and the expectation value converges for
both times to infinity to $+1$. For the initial maximally entangled
Bell states $\phi^\pm$ ($r_2=r_3=0$) the oscillation goes with the
sum of the times, different from the maximally entangled Bell states
$\psi^\pm$ ($r_1=r_4=0$) where the oscillation only depends on the
difference of the times. Thus for $\phi^\pm$ a violation of the Bell
inequality would occur earlier. However, it turns out that for no
maximally entangled state a violation can be found by numerically
optimizing with different standard methods (none guarantees a global
maximum).

For all phases $\phi_i=0$ we find the value $$S=2.1175$$ (state
$\xi$ ($r_1=-0.8335; r_2=r_3=-0.2446; r_4=0.4308$): $t_1=t_2=0;
t_3=t_4=5.77\tau_S$). If we also vary over the phases we obtain a
slightly higher value $$S=2.1596$$ (state $\chi$ ($r_1=-0.7823;
r_2=r_3=0.1460; r_4=0.5877; \phi_1=-0.2751; \phi_2=\phi_3=-0.6784;
\phi_4=0$): $t_1=t_2=1.79\tau_S; t_3=t_4=0$), see also
Fig.~\ref{CversusmixedS}~(a). For the above cases the concurrence
gives $$\mathcal{C}(\xi)=0.84 \,e^{-\Gamma (t_l+t_r)}\quad
\textrm{and}\quad \mathcal{C}(\chi)=0.94\, e^{-\Gamma
(t_l+t_r)}\,.$$

In Fig.~\ref{CversusmixedS}~(b)-(d) a purity versus concurrence
diagrams are drawn. For $\phi^+$ we notice that the ``decoherence''
caused by the decay exceeds the purity--concurrence values of Werner
states, which represent an upper limit for all possible decoherence
modes in this picture given by a Lindblad equation for an initially
maximally
entangled qubit state, Ref.~\cite{BDH,ZB}. %The Bell states $\psi^\pm$ have an even
%smaller decrease of concurrence than $\phi^+$ (???).
An early decay of one kaon, Fig.~\ref{CversusmixedS}~(c), exceeds
even the purity--concurrence value of maximally entangled mixed
bipartite qubit states (MEMS) \cite{MEMS}.

To sum up, the initial entanglement decreases with a sum of times,
and it goes first hand in hand with a decrease in purity which can
then for latter times increase again. For non--maximal entangled
state the decrease of purity is much faster than for the maximally
entangled states. This seem to help to violate the Bell--CHSH
inequality though the ratio of oscillation to decay is low.

All other meson systems have the same decay rate for both
mass--eigenstates, but no \textit{active} measurements are possible
due to their fast decay, a necessary condition for any test of local
realistic theories versus QM. For $B$-mesons the symmetric Bell
state $\psi^+$ violates formally the Bell inequality while $\psi^-$
does not, though both states have the same purity--concurrence
behavior. The violation of a Bell inequality depends strongly on the
parameters describing these systems, rather than on the amount of
entanglement.

%For all other meson systems the ratio oscillation to decay is also
%too low, except for the $B_s$--mesons where the ratio is $>19.9$
%(see Ref.~\cite{SBBGH}). However, also for non--maximally entangled
%states we find no violation. Different to all other meson systems,
%neutral kaons are special, they have quite different decay constants
%(decrease in entanglement goes only halve as fast) and due to their
%``long'' lifetime (few centimeters) \textit{active} measurements are
%possible, a necessary condition for any test of local realistic theories versus QM.\\

\section{Conclusions} We show how to treat a single and bipartite
decaying neutral kaon system in quantum mechanics and analyze the
properties of the corresponding states via purity, entanglement and
nonlocality. Only two degrees of freedom at a certain time can be
measured reducing the set of observables and leaving some elements
of the state undefined.

Different from photons, nonlocality is for the neutral kaon system a
quite ``dynamical'' concept as correlations of states evolving up to
different times are involved. For entangled photons there is no
difference whether in principle the correlations are measured after
one or several meters. With each measurement the experimenter
chooses among two observables: the quasi--spin and the detection
time. Consequently, considering Bell inequalities for mesons,
Eq.~(\ref{chsh}), one can vary in the quasi--spin space or vary the
detection times or both. If varying in the quasi--spins space and
for simplicity choosing all times equal to zero, it has been shown
in Ref.~\cite{BGH3} that there is a connection between nonlocality
and the violation of a symmetry in high energy physics, i.e. the
${\cal CP}$ symmetry ($\cal{C}$=charge conjugation,
$\cal{P}$=parity).

In this work we have discussed the choice of measuring on both sides
an antikaon versus no antikaon at a certain time, which can
experimentally be realized via inserting a piece of matter at a
certain position from the source (corresponding to the detection
time). We find a novel violation of the Bell--CHSH inequality for
certain initial states, which are more ``robust against
decoherence'' caused by the decay mechanism. For these states
---currently not available by experiments--- the concurrence is not
maximal in partial agreement with Ref.~\cite{GisinPRL2005,Scarani}
that optimal Bell tests do not require maximally entangled states
for systems with more than $2$ degrees of freedom. A higher amount
of entanglement doesn't necessarily imply an increase of a violation
of the Bell inequality under investigation, in fact the Bell
inequality need not to be violated at all.

Therefore, these results suggest that for the neutral kaon system
nonlocality and entanglement are indeed some distinct quantum
features which manifest themselves in a way different than that for
bipartite qubit or qutrit systems, and their relation is subtler
than
one naively expects.\\
\\
%\begin{acknowledgement}
\noindent\textit{Acknowledgement:} I want to thank N.~Gisin for the
invitation to Geneva. And I want to thank V.~Scarani for bringing
the non-local machines to my attention. EURIDICE HPRN-CT-2002-00311.
\\
\noindent\textit{Note added in proof:} As recently found by
V.~Scarani and myself the $\phi^-$ violates the Bell inequality
slightly, this will be published in a common work Ref.~\cite{HS1}.
%\end{acknowledgement}

\end{document}